# Experimental Characterization of Additively Manufactured Metallic Alloys for Electric Propulsion Applications

J. Chamberlain, A. Shashurin

*School of Aeronautics and Astronautics, Purdue University, West Lafayette, IN, USA*

**This work explores the sputtering of additively manufactured (AM) materials for use in gridded ion source applications. The first part of the paper uses 316L stainless steel as an example to demonstrate that additively manufactured material does not exhibit adverse sputtering behavior, such as higher sputter yield, compared with conventionally manufactured material. To this end, three additively manufactured 316L stainless steel samples were exposed to the beam of a KDC-40 electrostatic gridded ion source at three distinct energy levels of 400, 600, and 800 eV on each side of the sample for a duration of one hour. The samples were masked to create a distinct boundary between treated and untreated regions, identifiable using profilometry, and were biased to -18V for testing. Samples were then examined using a Bruker optical profilometer and further processed using the open-source software Gwyddion to evaluate the sputtering yield. The sputter yield varied in the range 0.2-2 atoms/ion for 400-800 eV ions and increased with ion energy. The measured sputtering yield was fairly consistent with predictions from analytical models developed in prior literature, while exhibiting some variations potentially due to added effects of increased temperature and oxide layers. The second part of the paper demonstrates feasibility of using an additively manufactured tungsten-rhenium grids in existing KDC-40 electrostatic gridded ion source.**

# 1. Introduction

Sputtering is a process in which an energetic particle bombards a solid, removing atoms from the near surface layers due to the energetic collision [1], such as the case of ejected ions impinging on the accelerator grid of a gridded ion source (GIS). While this process can cause undesired deterioration of components, it is also useful for the likes of surface cleaning, etching, and thin film deposition. Gridded ion sources see many applications across a variety of industries. They are frequently used in material processing where the ejected plasma can be applied for thin-film deposition, plasma etching, sputtering, and other surface modifications. Electric propulsion space applications of GIS are of special interest in this study, where they are used for maneuvering or station-keeping on satellites, or for deep space missions as the primary means of thrust.

Gridded ion sources make use of grids to generate electric potential distribution that accelerate ions through the gaps and eject them. In electric propulsion applications, the aforementioned surface modifications end up being an undesired consequence as a result of long expected lifetimes on the order of tens of thousands of hours where regular maintenance is not an option. For GIS, the accelerator grid (AG) is particularly prone to erosion, as its highly negative bias attracts the positively charged ions it is ejecting. This can be alleviated by installing a third grounded decelerator grid (DG), which reduces overall wear on the AG but changes its wear pattern [2]. The ions no longer impinge centrally between the apertures of the AG with a DG installed but instead wear away the edges of the aperture to create a rounded corner. This change in erosion affects the electric field near the aperture, which in turn affects the flow through the aperture. [2] This sputtering can be exacerbated by trying to pull too much ion current through the apertures, causing faster erosion and potentially warping the grids due to heating effects. [2]

A potentially useful solution to help alleviate some of the issues with sputtering in GIS is additive manufacturing. Additive manufacturing provides many benefits, such as the ability to construct more complex geometry, and can do it in fewer operations than conventional manufacturing methods. A common method of metallic additive manufacturing called laser powder bed fusion can manufacture components consistently, regardless of grid thickness, with no cusps that can occur with photochemical machining. Tungsten is an appealing option for both sputter resistance and additive manufacturing but comes a few issues. It is highly prone to cracking during the printing process, but this can be mitigated with print settings and alloying with other refractory metals such as rhenium. The alloying can cause mechanical properties to vary considerably depending on these same inputs, and there is usually little data to understand these. [3] However, it is possible to tailor raw materials and alloys to specific needs for additive manufacturing with some testing and estimates. This enables creation of alloys specifically tuned to resist sputter mechanisms for the required application. For other plasma-facing components in electric propulsion applications, one can tailor the surface finish with print settings, which can be used to affect sputter erosion. [4] This would be very valuable for components where a precision finish is not necessary.

Therefore, with the growth of metallic AM in space applications, it is worth studying the sputter characteristics of materials used to better understand their applicability in harsh environments. In this work, we first characterize the sputter yield of additively manufactured 316L stainless steel samples to assess any potential differences in sputter yield compared to conventionally manufactured material. In the second part of the paper, we demonstrate feasibility of integrating additively manufactured grids (composition of 76% Tungsten, 24% Rhenium) into a Kaufmann-Robinson KDC-40 gridded ion source as a replacement for nominal components.

## 2. Methodologies & Equipment

The sputtering was performed in a 0.652 $m^3$ vacuum chamber using a Kaufmann-Robinson KDC-40 gridded ion source with an LFN-2000 neutralizer attached. The test stand was built off of a ¼-20 breadboard on tracks inside the chamber and placed at $d$=15 or 30 cm from the face of the ion source. A 0.5 mm thick molybdenum mask with a 2.4 mm hole that would allow ions to strike the sample was placed in front of the stand. The sample was placed up against the mask to only expose the desired area, and then the complete assembly was biased to -18V with electrical feedthroughs to the chamber to ensure collection of ion saturation current by the sample. The assembly was electrically isolated from the breadboard/chamber with fiberglass posts. A diagram of the experimental setup is shown in **Figure 1**.

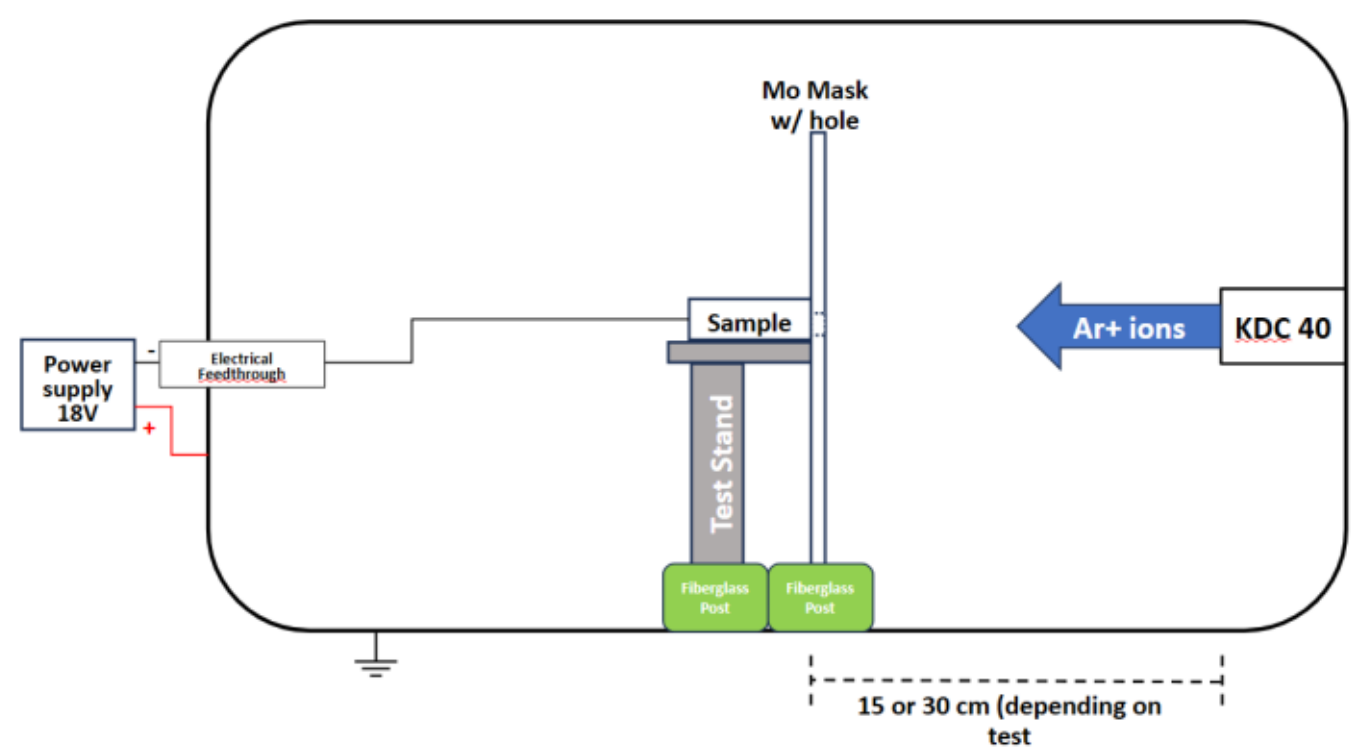

**Figure 1. Schematics of the experimental system utilized in this work**

The 316L stainless steel samples used in this work were made with a Colibrium Additive M2 laser powder bed fusion (LPBF) printer that uses selective laser melting (SLM). These samples were small blocks material measuring approximately 4.7 x 11.5 x 23 mm. The samples received the same heat treatment and other post-processing procedures that the grids did and had the same rough surface finish. To ensure the samples were able to be read on all sides, all surfaces were sanded procedurally with grits from 400 to 1200. After sanding, the samples were moved to the vacuum chamber for testing. The long, thin sides (11.5 x 4.7 $mm^2$) are coplanar with the original build plate that the block was manufactured from. The remaining smaller (4.7 x 11.5 $mm^2$) and larger sides (11.5 x 23 $mm^2$) are in the build direction. An example of the distinguishment between sides is shown in **Figure 2**.

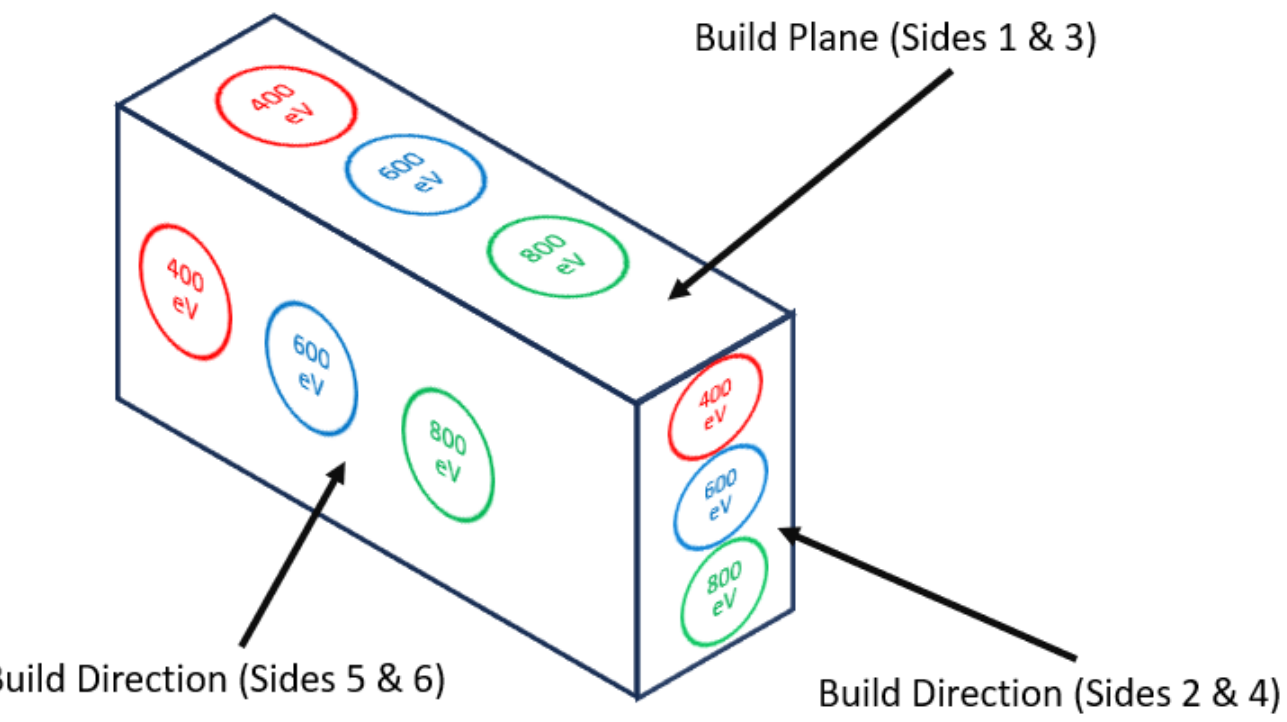

**Figure 2. Diagram of individual sample showcasing test layout for each side and side descriptors**

All samples after testing were examined under a Bruker GT-K optical profilometer. Etch rate ($R$) was characterized by measuring the sputter step height ($\Delta h$) using the profilometer and known test time ($\Delta t$) as $R=\Delta h/\Delta t$, and was later utilized for evaluating sputter yield. Each test site was imaged multiple times, and the presented results of each test site are averages of the multiple images for that respective region. Each test site uses at least two images to create an average value for its single data point. These images were analyzed using the open-source software Gwyddion, and the mean profile of each individual image was used to calculate the step height.

Ion beam current density at the sample location ($j_b$) was measured using a Langmuir probe (0.09689 cm$^2$ area) biased to -18 V and was later utilized for evaluating sputter yield. These measurements were performed for the duration of an entire testing cycle of one hour and then the current value was averaged over that timespan. In some tests, the neutralizer was manually toggled off to gather data about all possible operating conditions that represented what occurred during testing.

Sputter yield ($Y$) was determined as follows:

$$Y=\frac{R\rho_t e}{j_b m_t D} \quad (1)$$

where $m_t$ is the atomic mass of the target, $D$ is the atomic mass constant, $\rho_t$ is the density of the target, and $e$ is the elementary charge. The target density and atomic mass values must be the weighted average if the target is a compound material and the actual values are not measured or otherwise known. The density used here was calculated using a simple mass over volume as those quantities were known and measured ($\rho_t$=8000 $kg/m^3$). The atomic mass was calculated using a weighted average of the component materials using the compositional assumptions laid out in Section 3.1 ($m_t$=56.515$amu$).

# 3. Results and Discussion

## 3.1 Sputter Yield Measurements

The ion beam current density at the sample location ($j_b$) measured using a Langmuir probe is shown vs. beam energy ($E_b$) in **Figure 3** for an operational neutralizer at both distances. One can see the effects of the neutralizer being disabled lessening both with distance and higher beam energy.

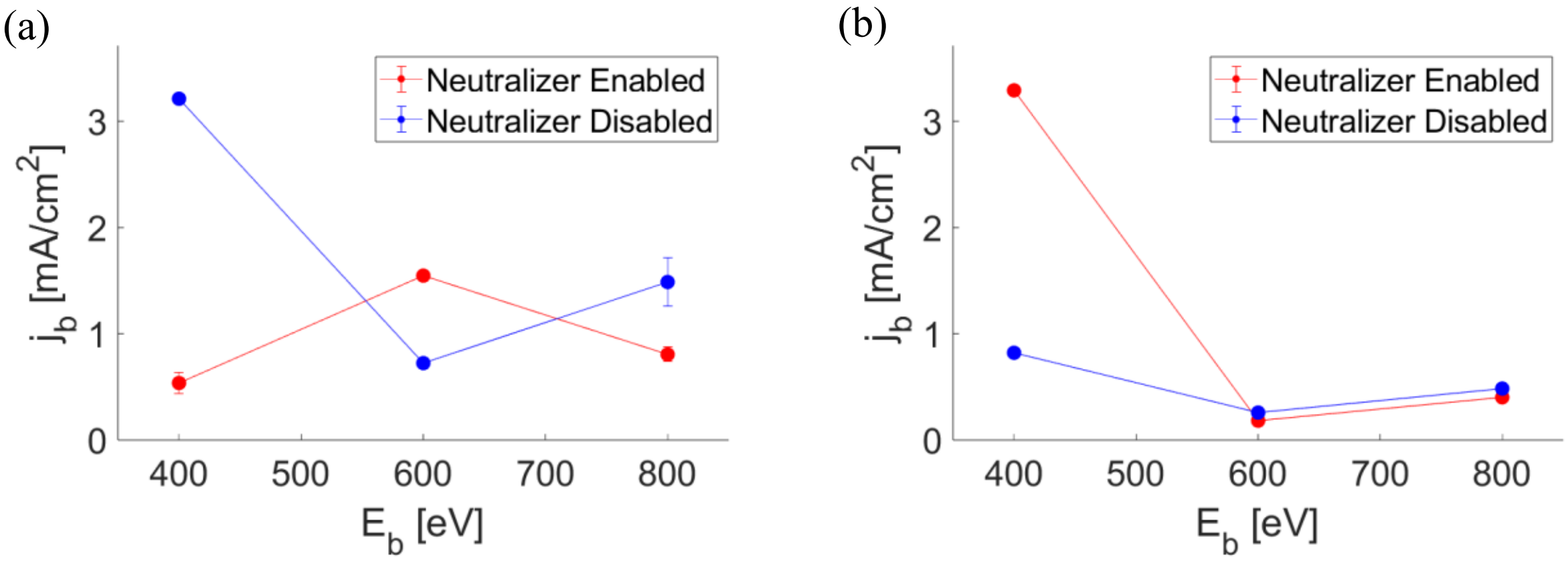

**Figure 3. Beam current ($j_b$) vs. beam energy ($E_b$) for testing distance (a) $d$=15 cm and (b) $d$=30 cm.**

A typical profile measured by the GT-K profilometer is shown in **Figure 4** for an 800 eV ion beam energy at the 30 cm distance. The test site images were analyzed using the open-source software Gwyddion, and the mean profile of each individual image was used to calculate the step height. The s-parameter in this fitting represents the vertical difference between the top and bottom plateaus, which is selected as the vertical step height of the profile ($\Delta h$). The majority of data points presented in **Figure 5** have standard errors around or below 15% for $\Delta h$, though some higher outliers exist due to low-quality surface conditions on test sites.

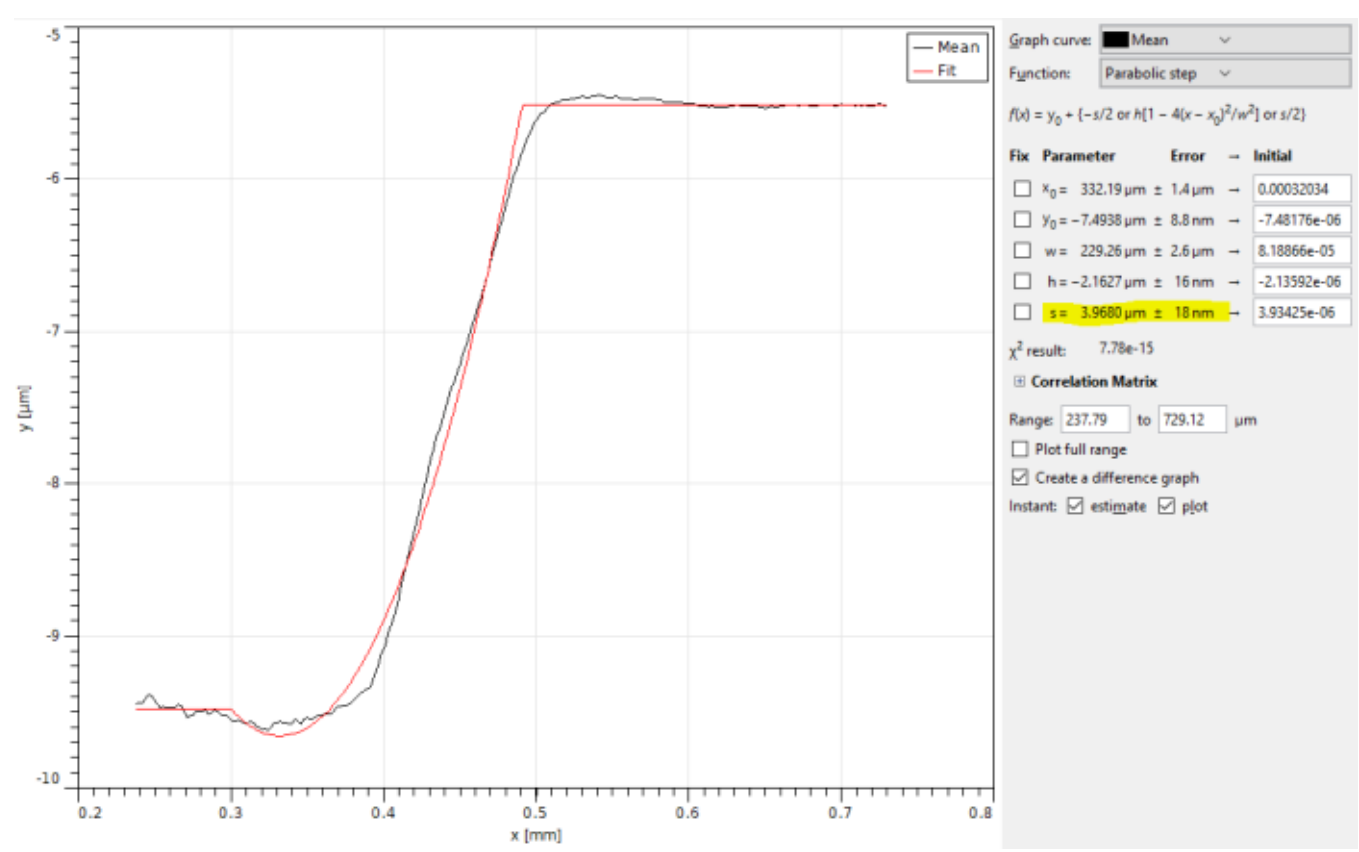

**Figure 4. A profile measured by the GT-K profilometer ($E_b$=800 eV, $d$=30 cm) and line fitting window of Gwyddion.**

Sputter yield results are presented in **Figure 5** for the build direction (a) and build plane (b). The individual data points are themselves an average of the scans for their respective sites, with uncertainties calculated through uncertainty propagation of Eq. (1). The results are presented alongside an uncertainty weighted average of every data point across all sides and samples. Additionally, Fig 5 outlines theoretical yields for elemental iron and a weighted average sputter yield of the components of 316L stainless steel, using a model developed Yamamura and Tawara. [5] These predictions are made using the Yamamura model due to its accuracy and the lack of 316L sputtering data at these specific energy levels with argon. Elemental iron being the bulk material in stainless steel may be a valid approximation of sputter characteristics. The other weighted average prediction for 316L is calculated using the primary components in 316L stainless at their maximum allowed concentrations according to the ASTM A240 standard: 0.03% C, 2.00% Mn, 0.75% Si, 18.0% Cr, 14.0% Ni, 3.00% Mo, and the remainder being Fe. The trace amounts of P, S, and N are excluded due to the lack of available sputter data and presumed minimal impact on sputter characteristics. Total sputter yield for multicomponent materials has been found to be typically quite different from a superposition of the yields of the components, but a mass fraction estimate can be used to get a rough estimate of expected yields. [1][6]

(a)

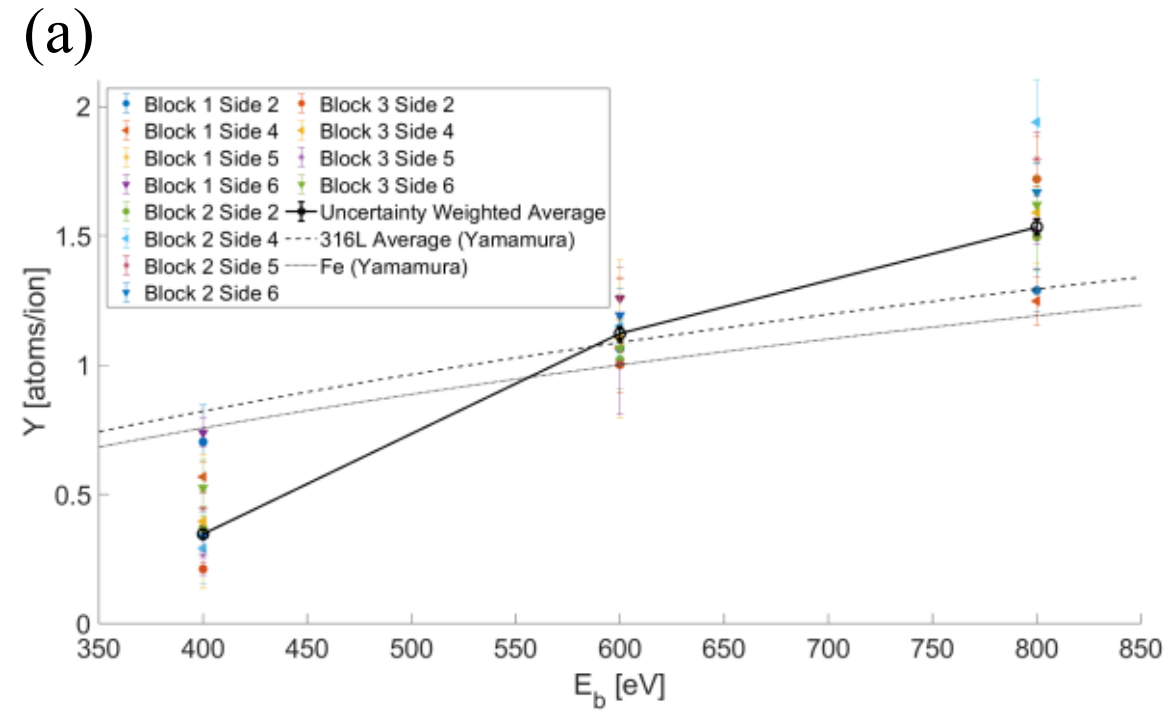

(b)

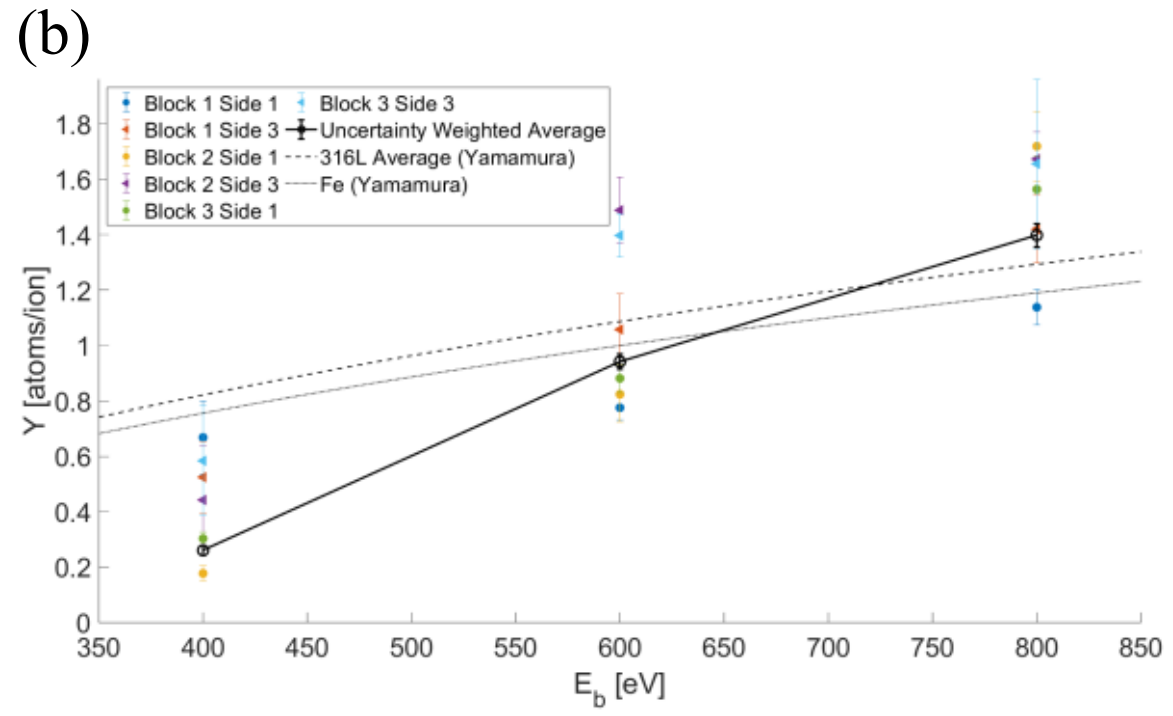

**Figure 5. Sputter yield measurements for (a) the sides in the build direction (sides 2, 4, 5, 6) and (b) the build plane (sides 1, 3) across all samples.**

One can see in **Figure 5**(a) that for the build direction sputter yield increases as the beam energy increases. The majority of sputter yield measurements at 400 eV trend lower than either prediction value at that energy level, primarily ranging between 0.21 and 0.73 atoms/ion. As energy increases, it becomes more in line with the predictions at 600 eV, but starting to trend above at 1.00 to 1.26 atoms/ion. By 800 eV the sputter yield exceeds either prediction significantly, mostly ranging from 1.49 to 1.94 atoms/ion. The variance weighted average for each energy level across a variety of test sites, indicated with the solid line in **Figure 5**(a), was fairly consistent with prior empirical data and analytical models, with 400 eV data trending below those predictions. Something that may not be readily apparent from the large amount of data presented is that a few corresponding sides match up fairly well. For Block 1 in particular, the short sides (2 and 4) and flat sides (5 and 6) align closely between themselves for 600 and 800 eV. This may indicate similar local composition in these sides from the manufacturing process. It is possible all sides may exhibit some level of anisotropy from sputtering depending on processes performed, rather than the build plane versus build direction traditionally expected with AM materials.

The sputter yield measurements in the build plane are shown in **Figure 5**(b). A similar trend is indicated to the build direction sides, where 400 eV is beneath the prediction, while it grows to match it at higher energy levels. The 600 eV energy level varies more considerably than the build direction, with one group from 0.82 to 0.88 atoms/ion and another two points clustered far above at 1.4 and 1.5 atoms/ion. The 800 eV points cluster slightly more evenly above the prediction as in the build direction, mostly ranging from 1.42 to 1.65 atoms/ion.

The observed lower sputter yield at 400 eV compared to the predicted values, as illustrated in **Figure 5**, can be explained by likely underestimation of the measured sputter yield at 400 eV, as the samples tested at this beam energy were associated with lower sample temperatures than those tested at higher beam energies. Specifically, while sample temperature was not controlled or monitored during these experiments, post-test evaluation of samples indicated significant increase in their temperature. This temperature increase is expected to have been minimal during the 400 eV tests, larger during the 600 eV tests, and largest during the 800 eV tests. At the same time, previous empirical results by Roth *et al* show an increase in sputtering yield of 316 stainless steel from room temperature up to 500 C. [7] Therefore, higher sputter yields at 400 eV can be expected if the sample-temperature effect is taken into account. Overall, additively manufactured 316L stainless steel exhibits fairly consistent sputter yield with that of traditionally manufactured material.

In this work, there were two main groups of factors governing the uncertainty of measured sputter yield. The first group was primarily driven by measurements errors of quantities such as $j_b$ and $R$. The combination of operational issues with the neutralizer and lack of in-situ $j_b$-data for individual tests likely provided considerable uncertainty for individual measurements. Also, the condition of the samples was likely a major impact on the $R$-data, as curvature of the sample surfaces caused issues when gathering data with the profilometer. The second group of factors governing the uncertainty of measured sputter yield was associated with a lack of full control over experimental conditions. One factor in this group is sample temperature, which was not controlled in these experiments. While there was no in-situ data for temperature of the samples, tests performed subsequently were noted to increase the samples' temperature (>260 C expected in some tests). A single typical day of testing involved the following sequence of tests: 400→600→800→400→600→800 eV. A 400 eV test could either follow up an 800 eV test that would significantly raise the temperature of the sample, or go from a cold start at the beginning of a new testing day, potentially increasing variation at that energy level. Another factor contributing to incomplete control over the experimental conditions was the presence of an oxide layer on the samples, since they were exposed to air, sometimes at elevated temperatures after testing. The presence of oxide layers is known factor to affect sputter yield. [7] [8] [9] [10]

The current work represents an initial study of the sputter yield in additively manufactured material, with a more comprehensive characterization planned for future work. Specifically, future work will utilize EDX to analyze the surface composition of the material for preferential sputter analysis before, during, and after sputtering. Grain and crystal structures of the material will be analyzed and compared to conventionally manufactured materials to explore any potential avenues that would affect energy transfer for cascade collisions. Given the rise in temperature seen during testing, a method for in-situ temperature control will be implemented. Finally, sputter yield in additively manufactured W-Re material will be measured.

### 3.2 Integration of AM Tungsten-Rhenium Grids

In this section we discuss utilization of AM W-Re grids (composition of 76% Tungsten, 24% Rhenium) in KDC-40 gridded ion source. The original KDC-40 grids were scanned and replicated with metal 3D printing (Figure 6), and then substituted in place of the original grids. As a result of changes in position of key features and form of the grids, several minor modifications were made to them to assemble the system. A fully assembled system with all modifications completed and the AM grids successfully integrated is shown in **Figure 7**.

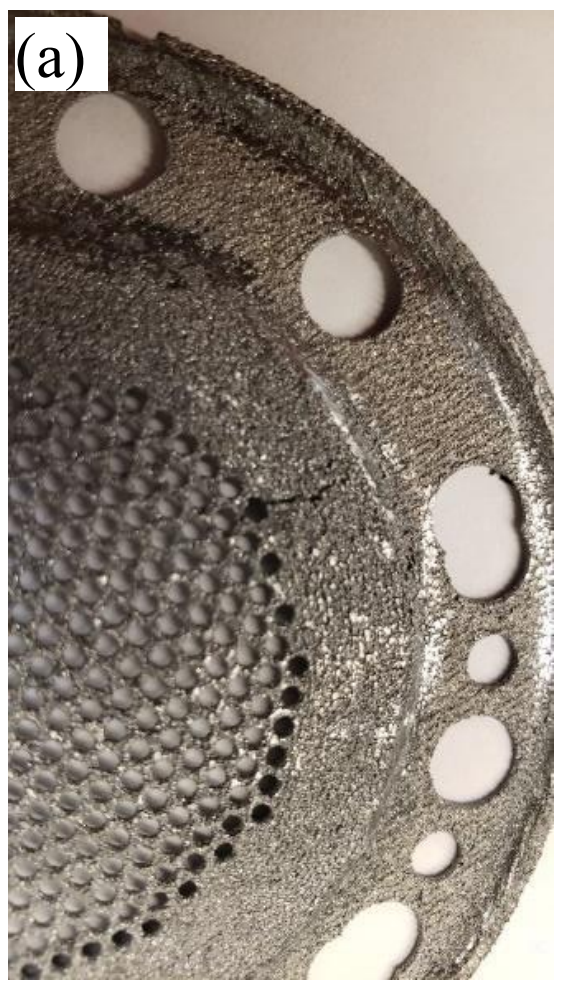

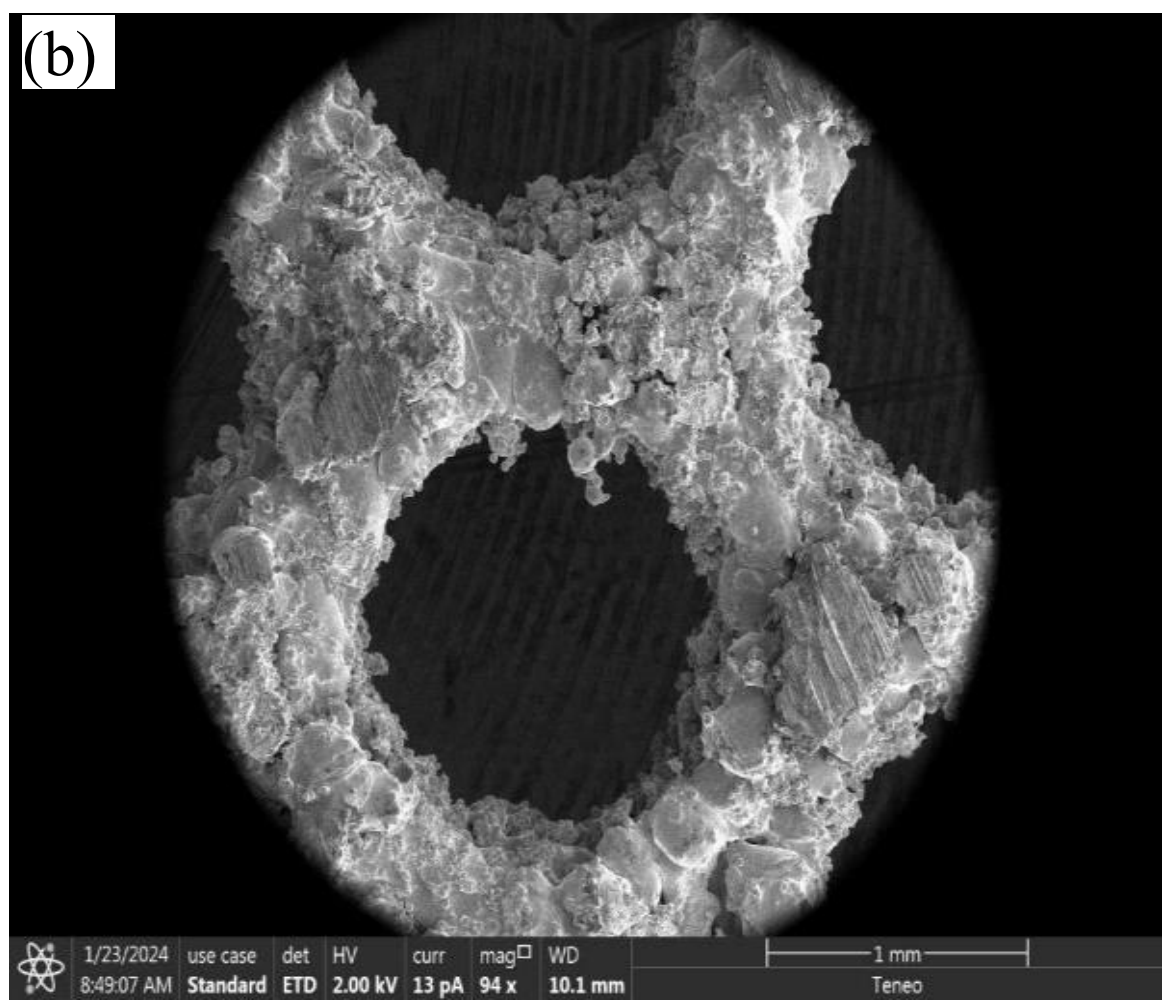

**Figure 6. A close up of the manufacturing defects on the (a) Decelerator Grid and (b) an SEM image of the Decelerator Grid**

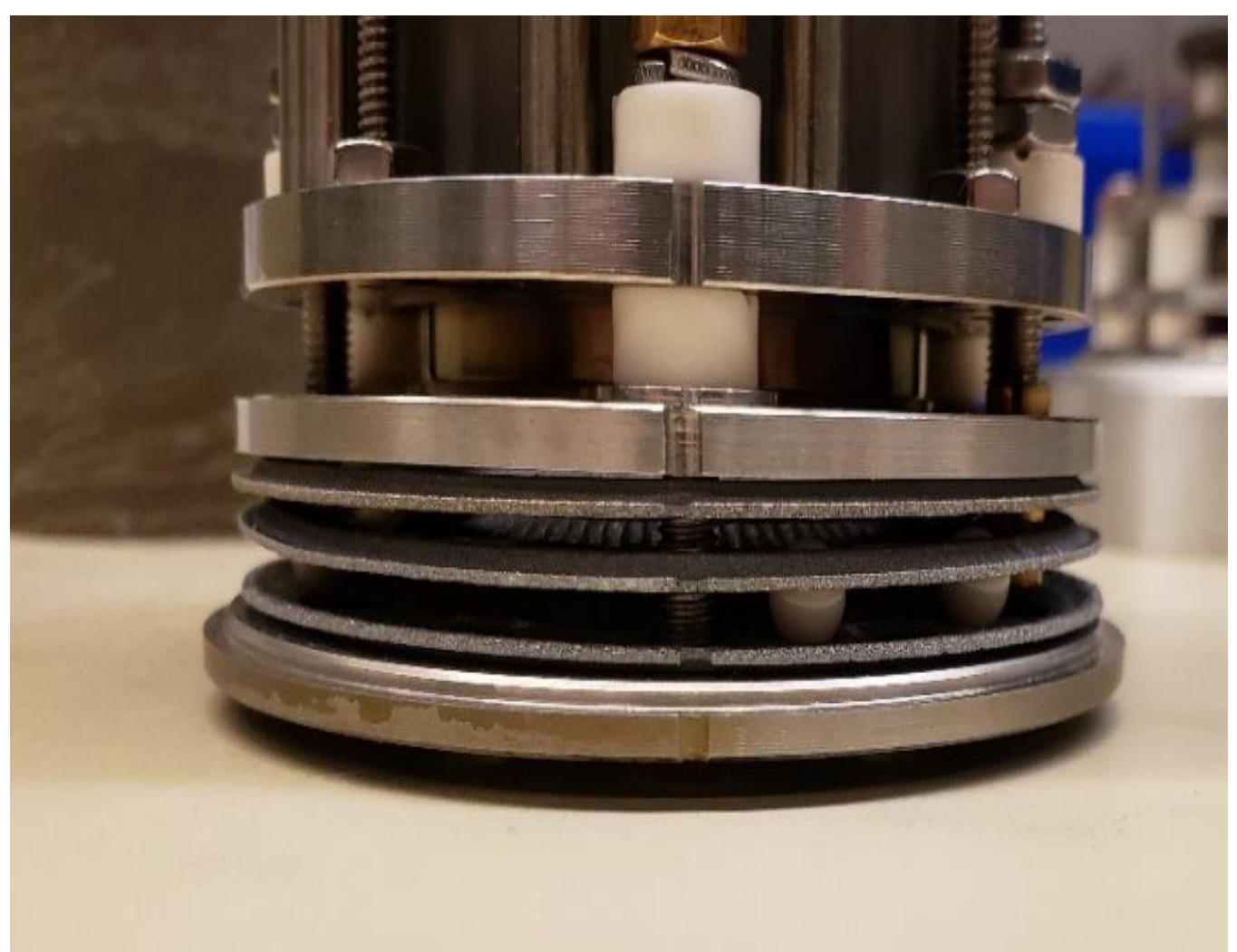

**Figure 7. Assembled Ion Optics using all three AM grids, largest ceramic spacers at each location, and raised electrical feedthroughs**

Testing was conducted over a series of experiments varying in duration, with a cumulative testing duration of one hour for the full system. The representative photograph of KDC-40 operation with three AM grids is shown in **Figure 8**. Stable, longer-term operation of the KDC-40 gridded ion source with AM W-Re grids was experimentally confirmed. The system appeared to have more stability issues at lower beam energy and would frequently cycle itself as the control stack attempted to reach a stable operational point.

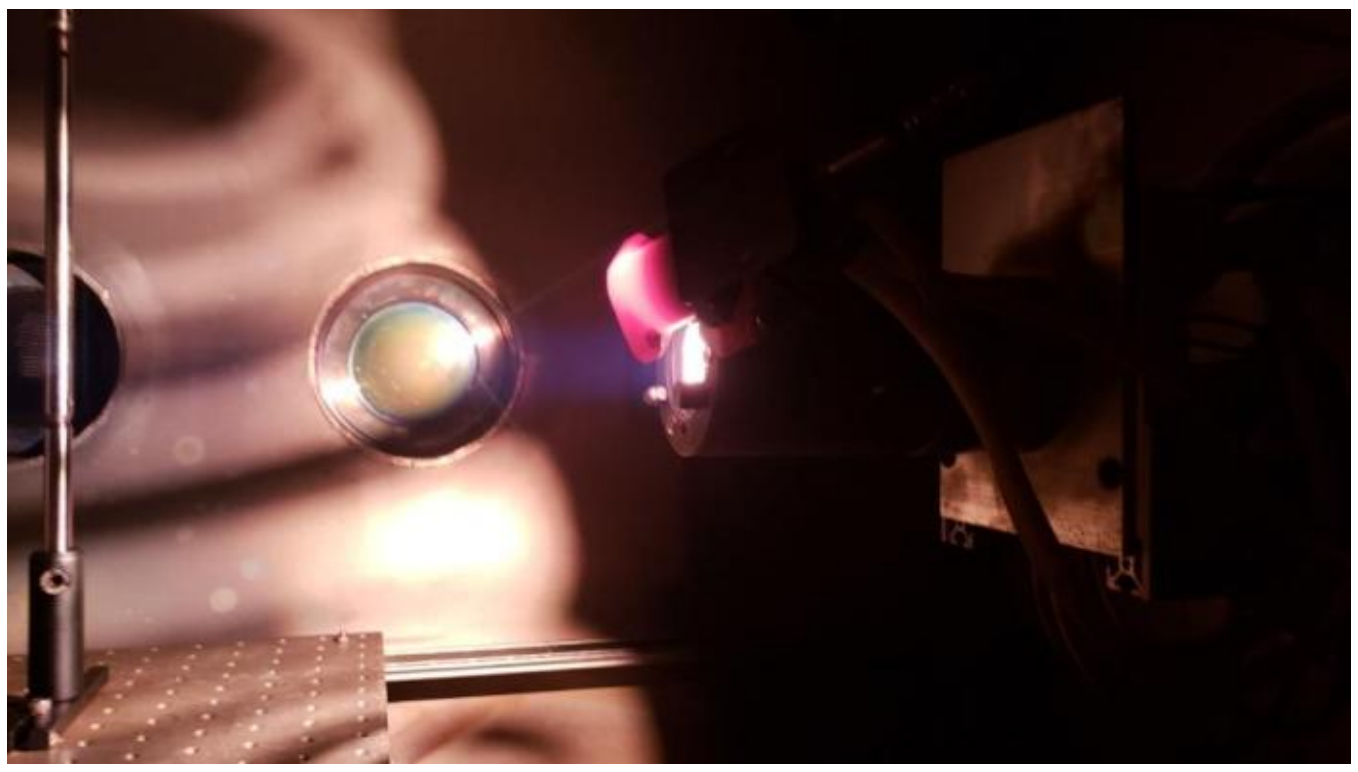

**Figure 8. Demonstration of stable operation of KDC-40 with three AM W-Re grids with clearly observed collimated ion beam.**

SEM images of the AM grids before testing with KDC-40 (pre-test images) and after operation (post-test images) were analyzed and compared, as shown in **Figure 9** (a) through (f). No substantial morphological changes of the AM grids were observed after the short series of tests conducted here. Note that longer duration tests would be required in order to accurately quantify erosion of the AM grids.

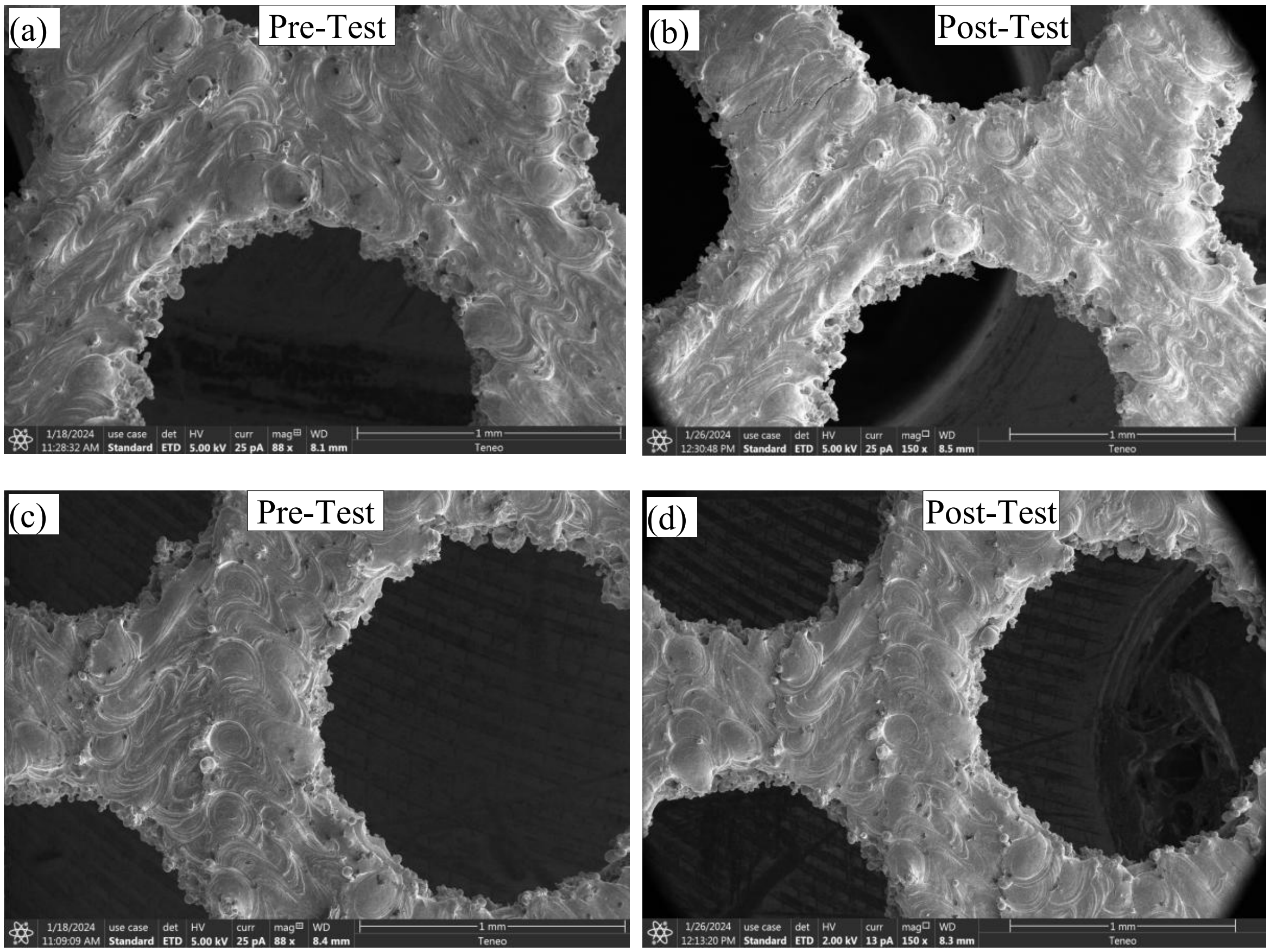

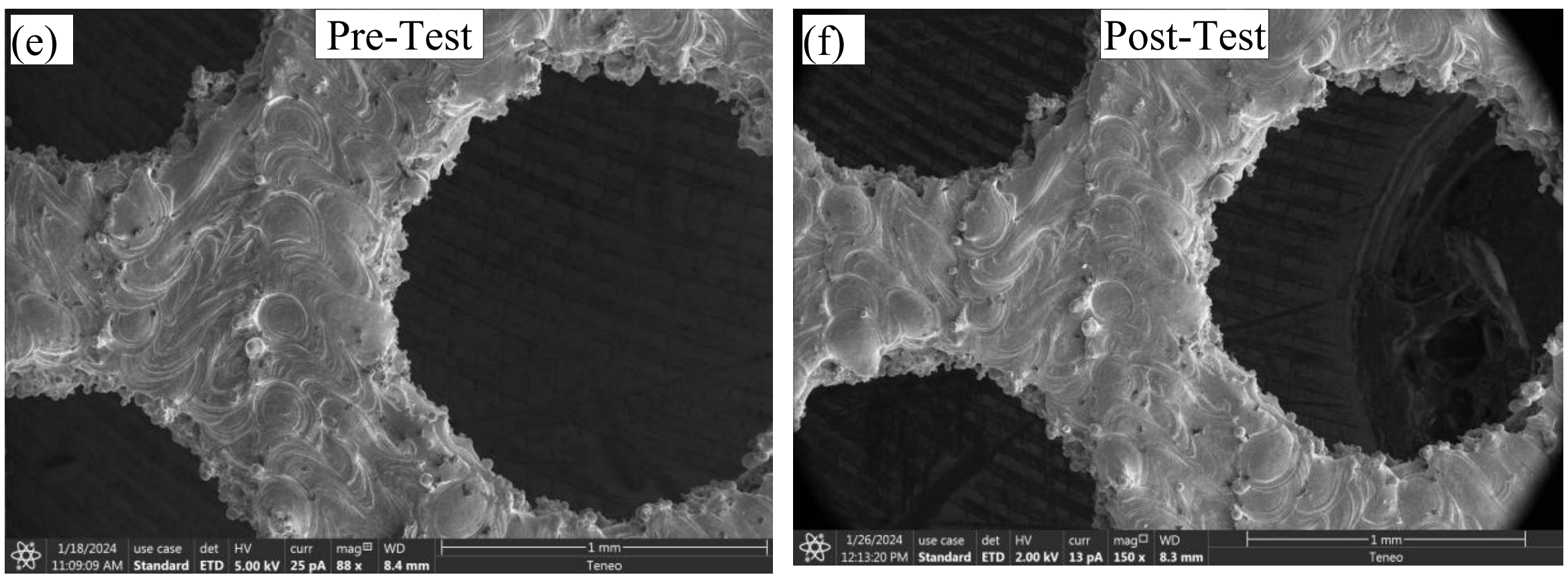

**Figure 9. Pre-test and Post-test SEM images of (a)-(b) SEM image of the Screen Grid, (c)-(d) Accelerator Grid, (e)-(f) Decelerator Grid**

## 4. Conclusions

To summarize, additively manufactured 316L stainless steel exhibits fairly consistent sputter yield with that of traditionally manufactured material. Observed variability in sputter yield are associated with the fact that oxide surface layers sputter more readily than their base element when considering total sputter yield, with additional variability occurring depending on evenness of the makeup of the underlying bulk material. Subsequent testing on the same sample also potentially adds the thermal effects to the base predictions of collisional sputtering, especially given the rise of sample temperature during testing due to efficient energy transfer between ion and target. Future work will require improved control over experimental uncertainties, such as sample temperature, air exposure, and sample flatness, as well as sputter yield measurements with additively manufactured W-Re material and long-duration testing of AM W-Re grids using a gridded ion source.

## ACKNOWLEDGMENTS

This project was supported NASA SBIR award #204084 to Quadrus Corporation. We thank Joseph Sims of Quadrus Corporation for providing the additively manufactured 316 stainless steel samples, W-Re grids, and useful discussions.

# REFERENCES

[1] Rainer. Behrisch and H. Henrik. Andersen, *Sputtering by Particle Bombardment I*, vol. 47. in Topics in Applied Physics, vol. 47. Berlin, Heidelberg: Springer Berlin Heidelberg, 1981. doi: 10.1007/3-540-10521-2.

[2] H. R. Kaufman, “Applications of Broad-Beam Ion Sources: An Introduction,” 2011.

[4] J. N. Brooks and D. N. Ruzic, “Modeling and analysis of surface roughness effects on sputtering, reflection and sputtered particle transport,” *Journal of Nuclear Materials*, vol. 176–177, pp. 278–282, Dec. 1990, doi: 10.1016/0022-3115(90)90060-Z.

[3] A. Talignani *et al.*, “A review on additive manufacturing of refractory tungsten and tungsten alloys,” *Additive Manufacturing*, vol. 58, p. 103009, Oct. 2022, doi: 10.1016/j.addma.2022.103009.

[5] Y. Yamamura and H. Tawara, “Energy dependence of ion-induced sputtering yields from monatomic solids at normal incidence,” *Atomic Data and Nuclear Data Tables*, vol. 62, no. 2, pp. 149–253, Mar. 1996, doi: 10.1006/adnd.1996.0005.

[6] G. Betz, “Alloy sputtering,” *Surf Sci*, vol. 92, no. 1, pp. 283–309, Feb. 1980, doi: 10.1016/0039-6028(80)90258-7.

[7] J. Roth, J. Bohdansky, W. O. Hofer, and J. Kirschner, “Erosion and changes in surface composition of stainless steel 316 after low energy light ion bombardment at temperatures between 50 and 660°C,” in *Proceedings of the International Symposium on Plasma Wall Interaction*, Elsevier, 1977, pp. 309–316. doi: 10.1016/B978-0-08-021989-9.50043-9.

[8] R. Kelly and N. Q. Lam, “The sputtering of oxides part i: a survey of the experimental results,” *Radiation Effects*, vol. 19, no. 1, pp. 39–48, Jan. 1973, doi: 10.1080/00337577308232213.

[9] R. Bastasz and G. J. Thomas, “Surface analysis of sputtered stainless steel,” *Journal of Nuclear Materials*, vol. 76–77, pp. 183–187, Sep. 1978, doi: 10.1016/0022-3115(78)90133-2.

[10] G. Betz and G. K. Wehner, “Sputtering of multicomponent materials,” 1983, pp. 11–90. doi: 10.1007/3-540-12593-0_2.